\documentclass[runningheads]{llncs}
\usepackage[T1]{fontenc}
\usepackage{graphicx}
\usepackage{enumitem}
\usepackage{hyperref}

\hypersetup{
    colorlinks=true,
    linkcolor=blue,
    filecolor=black,
    citecolor=blue,
    urlcolor=blue,
}

\begin{document}
\title{RiverBench: an Open RDF Streaming Benchmark Suite}
%
%
\author{Piotr Sowiński\inst{1,2}\orcidID{0000-0002-2543-9461} \and
Maria Ganzha\inst{1,2}\orcidID{0000-0001-7714-4844} \and
Marcin Paprzycki\inst{2}\orcidID{0000-0002-8069-2152}}
\authorrunning{P. Sowiński et al.}
%
\institute{Warsaw University of Technology, Warsaw, Poland\\
\email{piotr.sowinski.dokt@pw.edu.pl, maria.ganzha@pw.edu.pl} \and
Systems Research Institute, Polish Academy of Sciences, Warsaw, Poland
\email{\{firstname.surname\}@ibspan.waw.pl}}
\maketitle              
\begin{abstract}

RDF streaming has been explored by the Semantic Web community from many angles, resulting in multiple task formulations and streaming methods. However, for many existing formulations of the problem, reliably benchmarking streaming solutions has been challenging due to the lack of well-described and appropriately diverse benchmark datasets. Existing datasets and evaluations, except a few notable cases, suffer from unclear streaming task scopes, underspecified benchmarks, and errors in the data. To address these issues, we propose RiverBench, an open and collaborative RDF streaming benchmark suite. RiverBench leverages continuous, community-driven processes, established best practices (e.g., FAIR), and built-in quality guarantees. The suite distributes datasets in a common, accessible format, with clear documentation, licensing, and machine-readable metadata. The current release includes a diverse collection of non-synthetic datasets generated by the Semantic Web community, representing many applications of RDF data streaming, all major task formulations, and emerging RDF features (RDF-star). Finally, we present a list of research applications for the suite, demonstrating its versatility and value even beyond the realm of RDF streaming.

\end{abstract}

\section{Introduction}

RDF streaming applications are very diverse -- spanning many stream types~\cite{sowinski2023rdfstax}, application areas, and use cases. 
Benchmarking these solutions remains a major challenge -- although for some of the streaming tasks (e.g., streaming query engines) there are robust and well-established benchmarks, for others this is not the case. Especially acute is the lack of robust benchmarking datasets that would be sufficiently diverse, well-described, and useful in more than one benchmarking task. The existing benchmarks suffer from a lack of accessibility, poor descriptions (both human- and machine-readable), and unclear licensing.

To address these issues, in this work we introduce \textbf{RiverBench}, an open and collaborative RDF streaming benchmark suite that covers all major streaming task formulations (as per the RDF Stream Taxonomy -- RDF-STaX~\cite{sowinski2023rdfstax}), emerging RDF features (RDF-star), and is applicable even beyond the realm of RDF streaming. RiverBench addresses a set of requirements based on the observed deficiencies of current benchmarks, and best practices developed by the community. It aims to be a sustainable, community-driven hub for streaming RDF datasets.

\section{Background}
\label{sec:sota}

This section summarizes the current state of the art with regard to streaming RDF benchmarks. The aim of the summary is to provide a motivation for why RiverBench was created and for each of its design decisions.

Several RDF streaming benchmarks and datasets were proposed in the past, often tailored to a specific task formulation. In the streaming protocol-oriented research, the first benchmark was published with RDSZ~\cite{fernandez2014rdsz}, later expanded in the ERI paper~\cite{fernandez2014efficient}. The ERI benchmark\footnote{\url{http://rohub.linkeddata.es/RO-ISWC-14/}} consists of 16 datasets of various lengths and domains. The paper presents several statistics about the datasets, which helps with evaluating a method's performance on a given dataset. Each dataset is a flat N-Triples file (a flat RDF triple stream). In the evaluations the flat stream was grouped into RDF graphs of fixed size (e.g., 4096 triples), producing an RDF graph stream. However, this approach is often not representative of the use cases, as in reality element size can vary even message-to-message. Secondly, the benchmark is skewed towards some applications or ontologies, making it less representative of real workloads. For example, 5 of the datasets focus on weather measurements, and 3 of those (LOD\_*) follow the same schema. Thirdly, some of the datasets contain syntax errors~\cite{sowinski2022efficient}, which hampers their reuse. Finally, the license and provenance of each dataset is not clearly stated, making it hard to modify and republish them.

Many benchmark datasets and tools were developed for evaluating RSP reasoning and querying engines -- SRBench~\cite{zhang2012srbench}, LSBench~\cite{le2012linked}, CityBench~\cite{ali2015citybench}, RSPLab~\cite{tommasini2017rsplab}, WatDiv~\cite{gao2018stream}, Stream Reasoning Playground~\cite{schneider2022stream}, and more. These works have largely focused on evaluating the efficiency of query engines, and not on the datasets themselves, therefore, many have used data generators for a single type of stream. These generators tend to be tightly integrated with the specific frameworks or streaming models that are evaluated, and thus are very hard (if at all possible) to reuse outside their context. Secondly, synthetic datasets are a reasonable choice for some evaluations, but in others less so (such as compression efficiency evaluations). Here, notably, SRBench uses a real-world, streaming sensor dataset -- LinkedSensorData~\cite{patni2010linked}. CityBench also uses several real-world datasets from the CityPulse project~\cite{ali2015citybench}. The datasets of SRBench and CityBench are clearly licensed and can be obtained in standard RDF serializations. However, they are distributed as flat RDF triple streams, and thus lack an explicit split into stream elements.

\section{Requirements for Benchmark Datasets}
\label{sec:bench_proc}

Based on past research, observed issues in reproducibility, and established best practices (such as the Findability, Accessibility, Interoperability, and Reuse principles -- FAIR~\cite{wilkinson2016fair}), we propose a set of practical requirements for streaming RDF datasets to be used in benchmarks and other performance evaluations. These requirements are meant to ensure that the obtained results are representative of real-world performance of the method, and that they can be easily reproduced.

\begin{itemize}[leftmargin=0.8cm]
    \item[\textbf{R1}:] The datasets must be varied with regard to technical and non-technical aspects. The range of represented use cases, ontologies, and technical features should be as wide as possible. \\
    \textbf{Rationale:} The suite should be varied to be representative of a wide range of use cases, and to avoid skew towards any use case or technical aspect.

    \item[\textbf{R2}:] The datasets must be freely accessible and packaged in a common, easy-to-use format. \\
    \textbf{Rationale:} Ease of access is crucial to ensuring that the datasets will be systematically reused in future evaluations by researchers~\cite{wilkinson2016fair,frey2022reproducibility}.

    \item[\textbf{R3}:] The datasets must be well-described with documentation and machine-readable metadata. All relevant information about the technical and non-technical characteristics of the dataset must be included. \\
    \textbf{Rationale:} Information about the dataset's use case, technical characteristics, and more are often crucial to understanding why a method performs well or poorly with a given dataset~\cite{fernandez2014efficient}. Machine-readable metadata can also help automate some evaluation tasks, or serve as a source of information for research on the datasets themselves.

    \item[\textbf{R4}:] The license of the dataset and its metadata must be free and explicitly stated. The attribution of the source must be clear. The license must not prohibit commercial use or modification. \\
    \textbf{Rationale:} Clear licensing ensures that the datasets can be freely reused in evaluations. The licenses must be permissive so that the datasets can be adapted, republished, and used in a wide variety of contexts.

    \item[\textbf{R5}:] The datasets must be semantically versioned and the past versions must be accessible under a persistent URL (PURL). \\
    \textbf{Rationale:} This is a crucial requirement for ensuring reproducibility, as different versions of datasets will produce different, incomparable results.
    
    \item[\textbf{R6}:] The datasets must be explicitly divided into discrete elements (if applicable to the given problem). \\
    \textbf{Rationale:} Previous RDF streaming benchmarks did not always specify the division into elements explicitly~\cite{fernandez2014efficient,fernandez2014rdsz}, which allowed different methods to use different element sizes, making the approaches difficult to compare. This requirement does not apply to flat RDF streams.
    
    \item[\textbf{R7}:] The length of stream datasets must be at least 10,000 elements. \\
    \textbf{Rationale:} Longer streams provide more reliable benchmark results, reducing the effects of random noise, cache \emph{warm-up}, and other factors.

    \item[\textbf{R8}:] The datasets must be valid RDF that can be parsed easily. \\
    \textbf{Rationale:} Past benchmark datasets sometimes included errors~\cite{sowinski2022efficient}, hampering their reuse. This requirement calls for \emph{valid RDF} specifically, but, if clearly marked as such in metadata, distributing datasets using non-standard RDF features should also be possible.
\end{itemize}

\section{Structure of RiverBench}
\label{sec:structure}

This section presents the overall structure of the RiverBench benchmark suite, which puts the above theory to practice. The suite from its inception has been envisioned as an open, community-driven, and collaborative project, that follows FAIR principles and other established best practices. These crucial aspects have influenced every part of the suite, as described in detail below.

\subsection{Datasets}
\label{sec:str_datasets}

\begin{figure}[htb]
\centerline{\includegraphics[width=10.5cm]{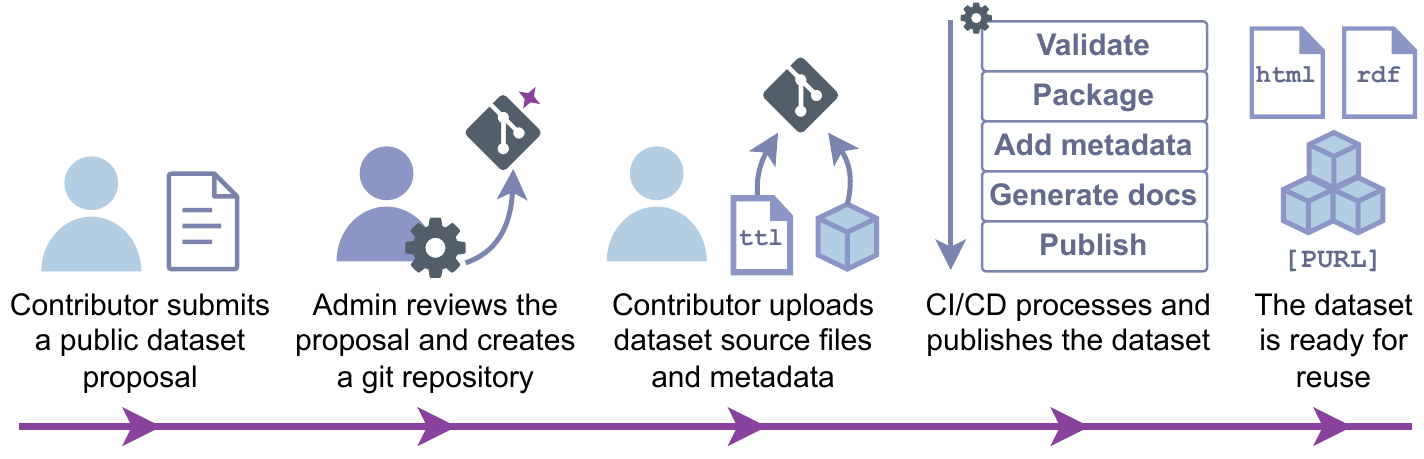}}
\caption{The process of adding and publishing a dataset in RiverBench.}
\label{fig:process}
\end{figure}

Datasets are added to RiverBench through a documented, public process that ensures that the relevant requirements are met (as proposed in the previous section). Firstly, the contributor fills out an issue template on GitHub, giving information about licensing, attribution, used stream type according to RDF-STaX, technical characteristics (e.g., stream length), the original use case, and more (\textbf{R1}, \textbf{R4}, \textbf{R6}, \textbf{R7}). An administrator reviews the template, asking for clarification if needed, validates that the requirements are met, and creates a public repository for the dataset. The contributor then uploads the dataset's source files and detailed metadata in a Turtle file. For both tasks, clear documentation and assistance is provided. The Turtle metadata file only includes information that must be specified manually (e.g., license, description, topics), and follows a predefined template. The rest of the process is fully automated, with Continuous Integration / Continuous Deployment (CI/CD) jobs checking the validity of the data, packaging the dataset, adding more metadata, generating documentation pages, and uploading the final packages. The dataset is validated with two RDF libraries (RDF4J and Apache Jena), using the strictest parsing settings (\textbf{R8}). Additional automatic checks ensure that the dataset's content is consistent with its metadata and the specified streaming task formulation. The full procedure of adding a new dataset is described in the documentation\footnote{\url{https://w3id.org/riverbench/documentation/creating-new-dataset}}. A visual summary of the process is presented in Figure~\ref{fig:process}. More details on the metadata, packaging, and publishing steps can be found in the following subsections.

By being completely public and open, the process is designed to invite external contributions from the community. It can also evolve over time, following community feedback (see section~\ref{sec:collaboration}). A key innovation over previous benchmarks is the inclusion of automated CI/CD jobs that guarantee that each dataset follows the same sequence of validation, packaging, and publishing steps.

\subsection{Benchmark Profiles}

RiverBench covers almost all stream types, as defined by RDF-STaX. However, in most applications, datasets representing different formulations should not be mixed. Thus, RiverBench employs a clear mechanism of grouping datasets into \emph{profiles}, with each profile corresponding to a specific task formulation and supported RDF features. The profiles also help research reproducibility, by being well-defined collections of datasets. Researchers can simply share the name and version (or the PURL) of the profile they used in the evaluation, to specify which datasets were involved. There are 7 base profiles in the current release of RiverBench, corresponding to different streaming task formulations:

\begin{itemize}
    \item \textbf{stream-subject-graphs} -- RDF subject graph streams.
    \item \textbf{stream-graphs} -- RDF graph streams, including subject graph streams.
    \item \textbf{stream-ts-named-graphs} -- timestamped RDF named graph streams.
    \item \textbf{stream-named-graphs} -- RDF named graph streams, including the timestamped variant.
    \item \textbf{stream-datasets} -- RDF dataset streams, including named graph streams.
    \item \textbf{stream-mixed} -- all grouped RDF streams, including graph and dataset streams.
    \item \textbf{flat-triples} -- flat RDF triple streams.
    \item \textbf{flat-quads} -- flat RDF quad streams.
    \item \textbf{flat-mixed} -- all flat RDF streams, including quad and triple streams.
\end{itemize}

Each of the base profiles has 4 variants that differ in the RDF features that are used in the datasets:

\begin{itemize}
    \item \textbf{(no suffix)} -- fully RDF 1.1-compliant datasets, with no extra features.
    \item \textbf{*-rdfstar} -- datasets may use RDF-star, but must remain compliant with the RDF-star draft specification~\cite{rdfstar}.
    \item \textbf{*-nonstandard} -- datasets may contain generalized triples and datasets, or other non-standard features, but not RDF-star.
    \item \textbf{*-rdfstar-nonstandard} -- datasets may use RDF-star and other non-standard features.
\end{itemize}

Thus, there are in total 36 profiles in the current version\footnote{\url{https://w3id.org/riverbench/profiles}}, which form a taxonomy that is reflected in the metadata and the documentation. This taxonomy not only helps organize the profiles, but also allows for making partial comparisons between evaluations made with different profiles. For example, if method A supports only streaming graphs and was evaluated with profile \texttt{stream-named-graphs}, it can be compared to method B that used the more general \texttt{stream-datasets-rdfstar} profile, as all datasets of the former are in the latter.

The conditions for datasets to be included in a profile are specified semantically in the profile's metadata file (in Turtle) and resolved by the CI automatically against the metadata of the datasets. Then, the CI generates the needed semantic relations and documentation. Creating a profile boils down to writing down its conditions, the rest is handled automatically. The profiles are also continuously updated whenever changes in datasets are made. The set of profiles can evolve over time to meet the community's needs, as described in section~\ref{sec:collaboration}.

\subsection{Metadata}
\label{sec:metadata}

Each dataset and profile in the suite has associated machine-readable RDF metadata describing its technical and non-technical aspects (\textbf{R3}). For interoperability, the metadata primarily uses the Data Catalog Vocabulary (DCAT) version 3~\cite{Cox:23:DCAT} and Dublin Core~\cite{dcterms}, with some additional properties from the FOAF and SPDX vocabularies. To represent the specific semantics of RiverBench, DCAT was extended with additional properties and classes. This additional ontology, although playing a secondary role in the suite, is fully documented and published under a PURL\footnote{\url{https://w3id.org/riverbench/schema/metadata}}, using best practices developed by the community~\cite{garijo2020best}, achieving 90\% score in \emph{FOOPS!}~\cite{garijo2021foops} 
and 4/4 stars in DBpedia Archivo~\cite{frey2020dbpedia}.

Part of the metadata is written manually, but a large portion is generated automatically during the packaging process. The auto-generated part includes rich metadata of the packages and dataset statistics. What follows is a high-level overview of the metadata that is included with each dataset.

\subsubsection{Non-technical Metadata}

\begin{itemize}
    \item \textbf{Description} -- information about the broader context in which the dataset was used or generated.
    \item \textbf{License and attribution} -- licensing and authorship information of the dataset, with a link to the license's definition in the SPDX License List~\cite{spdx_license}.
    \item \textbf{Topics / themes} -- tags selected from a pre-defined SKOS taxonomy, describing the dataset's application area and the type of included data.
\end{itemize}

\subsubsection{Technical Metadata}

\begin{itemize}
    \item \textbf{Used ontologies} -- a list of the most prominently used ontologies in the dataset.
    \item \textbf{Stream type} -- annotated using the RDF-STaX ontology~\cite{sowinski2023rdfstax}. Each dataset is assigned two stream types: one for the grouped stream formulation, and another for the flat.
    \item \textbf{Stream element split type} -- how was the dataset divided into stream elements. The types can be combined and include: \textit{statement count}, where the stream was split into elements of arbitrary length; \textit{topic}, where each element contains information about a different subject; \textit{time}, for elements associated with a particular point in time or a time interval. Additional details on the split can also be provided.
    \item \textbf{Temporal property IRI} -- used only for streams with a time-based element split, indicates the RDF property that marks the timestamps of stream elements. It is also the time property in timestamped RDF graph streams.
    \item \textbf{Stream element count} -- total number of stream elements in the dataset.
    \item \textbf{RDF features used} -- a checklist of major standard and non-standard RDF features used in the dataset.
    \item \textbf{Distributions (packages)} -- information about the available packaged variants of the dataset (see section~\ref{sec:packaging}), including byte size, checksums, content types, download URLs, and more.
    \item \textbf{Statistics} (for each distribution) -- automatically calculated statistical parameters (mean, standard deviation, minimum, maximum, sum) about several variables in the population of stream elements. The variables include the count of: IRIs, blank nodes, literals (simple, datatype, language), quoted triples (RDF-star), subjects, predicates, objects, graphs, and statements.
\end{itemize}

The metadata can be downloaded via links on the HTML documentation pages, as well as using the HTTP content negotiation mechanism on their PURLs (supported formats are: RDF/XML, N-Triples, Turtle, and the Jelly binary format~\cite{sowinski2022efficient}). 
The metadata is comprehensive, with the hope of enabling quantitative research on the effect of different types of datasets on systems processing RDF. It is released under a permissive license (CC BY 4.0) and can be freely reused for any purpose (\textbf{R4}). The metadata can be easily expanded, with dataset contributors being able to attach any additional semantic information to their datasets. The automatically generated information can also be extended with ease, for the entire suite.

\subsection{Packaging and Publishing}
\label{sec:packaging}

The datasets are packaged by the CI into files in a common format based on established standards (\textbf{R2}). The source archives uploaded by contributors are processed by the CI in a streaming manner, making the process scalable to even very large datasets. Each stream element is parsed, validated, and re-serialized. For every dataset three types of distributions are made: the grouped streaming distributions, the flat distributions (flat RDF streams), and the binary Jelly distributions, which can be used as both flat and grouped RDF streams. 
For each type, several stream length variants are prepared, starting with 10,000 elements, and increasing by a factor of 10 with each step. The length variants can be useful when testing methods that do not require very large datasets, or when streams of equal length are needed for a fair comparison.

In grouped RDF stream distributions, elements are serialized in W3C-standard Turtle or TriG, depending on the type of the stream. If the dataset uses RDF-star, the Turtle-star and TriG-star formats~\cite{rdfstar} are used. Stream elements are stored as separate, sequentially numbered files in a .tar.gz archive, which is a common, interoperable, and streamable format. The files in the archive are laid out sequentially, allowing the package to be processed in a streaming manner (one element at time, without decompressing the whole archive).
Flat RDF streams are simply serialized as either N-Triples or N-Quads files, or, in case RDF-star is used, N-Triples-star and N-Quads-star. The serialized data is gzip-compressed. This format is also streamable, as the file can be decompressed on-the-fly and read line-by-line.

These formats were chosen due to them being widely supported, ensuring the accessibility of the datasets. They were also specifically designed to be as simple as possible, while maintaining good performance characteristics. 
Additionally, the datasets are distributed in the high-performance Jelly format, which natively supports grouped and flat RDF streams. The Jelly distributions have the advantage of faster loading times, which can significantly speed up running benchmarks. Full documentation of the packaging formats can be found on the website\footnote{\url{https://w3id.org/riverbench/documentation/dataset-release-format/}}.

Packaged datasets are published along with their metadata and documentation under the RiverBench PURL. The releases use a versioning scheme derived from Semantic Versioning~\cite{semver}, giving clear compatibility guarantees between different releases and allowing users to refer to a specific stable version (\textbf{R5}). The profiles and the entire suite are also semantically versioned.

\subsection{Website and Documentation}

\begin{figure}[htb]
\centerline{\includegraphics[width=8.6cm]{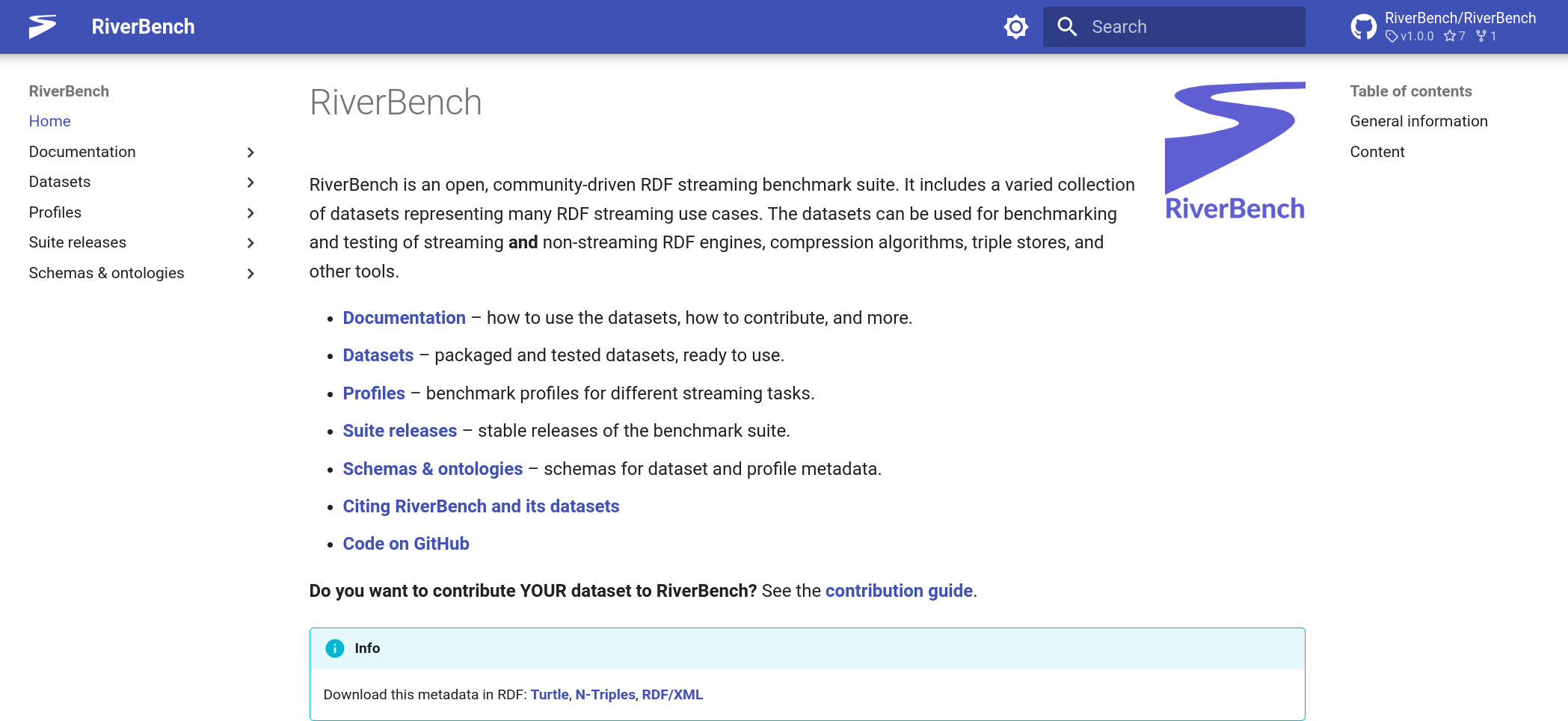}}
\caption{Screenshot of the main page of the RiverBench website.}
\label{fig:website}
\end{figure}

RiverBench's publicly available website gathers all information about the suite's datasets, profiles, documentation, and licensing (\textbf{R3}). Dataset and profile documentation is generated automatically from the metadata, ensuring that all information is human-readable and accessible. Documentation for older versions of datasets and profiles is preserved for future reference (\textbf{R5}). The website is hosted under a PURL registered with w3id.org. Under the same base PURL, machine-readable metadata for all resources is available (in RDF), as well as download links to dataset packages. Figure~\ref{fig:website} presents the main page of the website.

\subsection{FAIR Compliance -- Summary}

The following is a summary of how RiverBench complies with the FAIR principles~\cite{wilkinson2016fair}. The listed mechanisms are applied systematically and uniformly to every resource (dataset, profile, etc.) in RiverBench and enforced with pre-defined community processes and CI jobs.

\begin{itemize}[leftmargin=1.05cm]
    \item[\textbf{F1}:] All of RiverBench's resources are assigned a globally unique and persistent URL (under \url{https://w3id.org/riverbench}).
    \item[\textbf{F2}:] RiverBench's resources are described with rich metadata with links to external sources of information.
    \item[\textbf{F3}:] Metadata of resources clearly indicates the identifiers of the data, using appropriate RDF properties.
    \item[\textbf{F4}:] RiverBench's website is indexable by search engines and serves as a hub for accessing other resources. Additionally, RiverBench is archived in Zenodo, and its ontologies are registered in DBpedia Archivo~\cite{frey2020dbpedia} and Linked Open Vocabularies~\cite{vandenbussche2017linked}.
    \item[\textbf{A1}:] The resources are easily accessible via standardized protocols (HTTPS) and mechanisms (content negotiation).
    \item[\textbf{A2}:] The entirety of RiverBench, including its metadata, is archived in Zenodo. See also the sustainability plan in section~\ref{sec:collaboration}.
    \item[\textbf{I1}:] RiverBench uses RDF 1.1 and standard serializations for its metadata.
    \item[\textbf{I2}:] The metadata of resources uses established vocabularies (DCAT, Dublin Core, FOAF), as well as a custom ontology extension that follows the FAIR principles as well.
    \item[\textbf{I3}:] The resources in RiverBench are interlinked through well-defined links in their metadata. They also include references to external sources of information, such as SPDX licenses or ontologies used in datasets.
    \item[\textbf{R1}:] Dataset metadata includes rich contextual information about the context in which the dataset was created. The metadata of resources is plentiful, including, for example, detailed statistics about each dataset distribution.
    \item[\textbf{R1.1}:] All resources are clearly and permissively licensed.
    \item[\textbf{R1.2}:] All resources have human- and machine-readable authorship and provenance information associated with them.
    \item[\textbf{R1.3}:] RiverBench makes a best effort to implement best practices for metadata developed by the Semantic Web community, as presented above.
\end{itemize}

\section{Datasets Included in the Current Release}
\label{sec:datasets}

At the time of writing, the development version of RiverBench features 12 datasets\footnote{\url{https://w3id.org/riverbench/datasets}}, of which 9 are RDF triple streams, 2 are graph streams, and 1 is a quad stream. All datasets were added by the authors using the public community process described in section~\ref{sec:str_datasets}, which can be reviewed in the suite's issue tracker. The authors had direct involvement in only two of the datasets (from ASSIST-IoT), with the rest being gathered from community use cases.

\begin{itemize}
    \item \href{https://w3id.org/riverbench/datasets/assist-iot-weather/1.0.0}{assist-iot-weather} \& \href{https://w3id.org/riverbench/datasets/assist-iot-weather-graphs/1.0.0}{assist-iot-weather-graphs} -- two variants (an RDF graph stream and a timestamped RDF named graph stream) of the same source dataset -- SOSA/SSN weather measurements collected in the ASSIST-IoT project~\cite{szmeja2023assist}.
    \item \href{https://w3id.org/riverbench/datasets/citypulse-traffic/1.0.0}{citypulse-traffic} \& \href{https://w3id.org/riverbench/datasets/citypulse-traffic-graphs/1.0.0}{citypulse-traffic-graphs} -- two variants (an RDF graph stream and a timestamped RDF named graph stream) of the road traffic dataset from the CityPulse project, also used in CityBench~\cite{ali2015citybench}.
    \item \href{https://w3id.org/riverbench/datasets/dbpedia-live/1.0.0}{dbpedia-live} -- high-volume RDF graph stream from the DBpedia Live service, describing changes in Wikipedia~\cite{morsey2012dbpedia}.
    \item \href{https://w3id.org/riverbench/datasets/digital-agenda-indicators/1.0.0}{digital-agenda-indicators} -- RDF subject graph stream of very regular statistical information about the European information society~\cite{digital_agenda}.
    \item \href{https://w3id.org/riverbench/datasets/linked-spending/1.0.0}{linked-spending} -- RDF subject graph stream, part of the LinkedSpending dataset, which collects government spending statistics from around the world~\cite{hoffner2016linkedspending}.
    \item \href{https://w3id.org/riverbench/datasets/lod-katrina/1.0.0}{lod-katrina} -- RDF graph stream, the Katrina weather measurement dataset from LinkedSensorData~\cite{le2012linked}. Also used by SRBench~\cite{zhang2012srbench} and ERI~\cite{fernandez2014efficient}.
    \item \href{https://w3id.org/riverbench/datasets/muziekweb/1.0.0}{muziekweb} -- RDF subject graph stream, a Dutch knowledge base about music~\cite{muziekweb}.
    \item \href{https://w3id.org/riverbench/datasets/nanopubs/1.0.0}{nanopubs} -- RDF dataset stream, a set of freely-licensed Nanopublications, small units of publishable information~\cite{kuhn2018nanopublications}.
    \item \href{https://w3id.org/riverbench/datasets/politiquices/1.0.0}{politiquices} -- RDF graph stream describing news articles in Portuguese and the presented political stances~\cite{politi}.
    \item \href{https://w3id.org/riverbench/datasets/yago-annotated-facts/1.0.0}{yago-annotated-facts} -- RDF-star subject graph stream, a subset of YAGO 4~\cite{pellissier2020yago} including only fact annotations in RDF-star.
\end{itemize}

In the current release there is one RDF-star dataset, three datasets using RDF quads, and no datasets that would use non-standard RDF features beyond RDF-star. The main reason for this is that very few such datasets were publicly released and the authors had limited resources to prepare the datasets themselves. Nonetheless, RiverBench is to our knowledge the most technically diverse RDF streaming benchmark to date, and its collection of datasets is expected to expand in future releases with the help of the Semantic Web community. We have deliberately avoided adding more similar datasets to the suite (e.g., other datasets from LinkedSensorData or CityBench), to keep RiverBench as diverse as possible.

\section{Applications}
\label{sec:app}

RiverBench, by addresssing many task formulations, has broad application potential. This section presents the envisioned research applications of the suite.

\begin{itemize}
    \item Benchmarking streaming RDF protocols, such as ERI~\cite{fernandez2014efficient}, S-HDT~\cite{hasemann2012shdt}, and Jelly~\cite{sowinski2022efficient}. The \texttt{stream-*} profiles can be used for this purpose.
    \item Benchmarking RDF parsers and serializers (streaming and non-streaming), on a wide variety of datasets (including RDF-star).
    \item Evaluating compression performance of compact RDF representations, both streaming (ERI, S-HDT, Jelly, etc.) and non-streaming (e.g., HDT~\cite{fernandez2013binary}, with the \texttt{flat-*} profiles).
    \item Serving as a basis for streaming and non-streaming reasoning and querying benchmarks. For this, additional work of defining the queries and/or reasoning rules would be needed.
    \item Stress-testing RDF processing systems with representative, real-world data.
    \item Testing compatibility of RDF processing systems with real datasets that may contain uncommon edge cases (e.g., large literals, complex graphs).
    \item Conducting research on the streaming datasets, their complexity, applications, performance characteristics, etc.
\end{itemize}

More applications are expected to be identified over time, and RiverBench will welcome requests for enhancements that can help broaden its audience.

\section{Collaborative Development and Sustainability Plan}
\label{sec:collaboration}

To remain representative of the community's evolving needs, RiverBench must facilitate collaboration of various stakeholders from the industry and academia. Therefore, it takes a thoroughly open approach, with its every aspect being fully public and permissively licensed (metadata and documentation under CC BY 4.0, code under Apache 2.0). The main hub of activity is the RiverBench GitHub organization, which includes the source code and the issue tracker, freely accessible to any contributor. A comprehensive contributor's guide was prepared\footnote{\url{https://w3id.org/documentation/contribute/}}, which, along with detailed documentation, explain how to contribute new datasets, improve metadata and documentation, and modify the code.

\paragraph{Sustainability Plan.} Making RiverBench sustainable was one of the primary considerations from the start. RiverBench does not incur \emph{any} infrastructure maintenance costs, due to it being hosted exclusively on the permanently free and well-established infrastructure of GitHub and w3id.org. In case of, for example, a dramatic change to GitHub's open source project policy, the project requires for its basic needs only static file hosting and can be moved easily to a different service by the virtue of PURLs. RiverBench's source code and datasets are also archived independently in Zenodo. Regarding labor costs, the project aims to attract a wide audience, by being applicable in many scenarios. It is hoped that this will allow RiverBench to build a stable community that will be able to maintain it sustainably. Moreover, the vast majority of processes in the suite are fully automated, therefore, the workload needed to maintain it is minimal.

\section{Conclusion}
\label{sec:conclusions}

In this paper, we tackle the relevant issue of the lack of datasets for evaluating the very diverse RDF streaming solutions. We outline the requirements for streaming benchmark datasets, and introduce RiverBench, an open RDF streaming benchmark suite that applies these principles in practice. We hope that the suite, having the values of FAIR, community collaboration, and sustainability at its core, will be welcomed by the community as a useful resource. We aim to continue improving RiverBench in the future, especially by expanding its coverage of less-common streaming tasks, improving its metadata, and its tooling. The possible future enhancements can be discussed (and new ones can be proposed) in the project's issue tracker.

We invite all researchers and practitioners in the fields of Semantic Web and Knowledge Graphs to collaborate on future versions of RiverBench, to make it closely aligned with the community's needs. Especially welcome will be new datasets from varied application areas. We hope that with future community-led efforts, the suite will greatly improve in terms of quality, size, and diversity.

\paragraph*{Resource Availability Statement:}

\begin{itemize}
    \item Website of RiverBench: \url{https://w3id.org/riverbench}
    \item Source code for RiverBench and its tools: \url{https://github.com/RiverBench}
    \item Public issue tracker: \url{https://github.com/RiverBench/RiverBench/issues}
    \item Archive / backup in Zenodo: \url{https://doi.org/10.5281/zenodo.7909063}
\end{itemize}

\subsubsection*{Acknowledgements}

This work was partially funded from the EU’s Horizon 2020 research and innovation programme under grant agreement No 957258, in the ASSIST-IoT project.

\newpage

\bibliographystyle{splncs04}
\small{
\bibliography{bib/bibliography}
}
\end{document}